\documentclass{PoS}

\title{Isospin breaking corrections to the HVP at the physical point}

\ShortTitle{IB corrections to the HVP at the physical point}

\author{\speaker{Vera G\"ulpers}$^{1,2}$, Andreas J{\"u}ttner$^1$, Christoph 
Lehner$^3$, Antonin Portelli$^{2}$\\
        $^{1}$School of Physics and Astronomy, University of Southampton,
        Southampton SO17 1BJ, UK\\
        $^{2}$School of Physics and Astronomy, University of Edinburgh,
	Edinburgh EH9 3JZ, UK\\
	$^3$ Physics Department, Brookhaven National Laboratory, 
	Upton, NY 11973, USA
	\\
        E-mail: \email{Vera.Guelpers@ed.ac.uk}}

\author{RBC and UKQCD Collaborations}

\abstract{A determination of the hadronic vacuum polarization contribution to 
the anomalous magnetic 
moment of the muon from lattice QCD aiming at a precision of $1\%$ requires to 
include isospin breaking corrections in the 
computation. We present a lattice calculation of the QED and strong isospin 
breaking corrections to the hadronic vacuum polarization with Domain Wall 
fermions. The results are 
obtained using quark masses which are tuned such that pion and kaon masses agree 
with their physical values including isospin breaking corrections.}

\FullConference{The 36th Annual International Symposium on Lattice Field Theory - LATTICE2018\\
		22-28 July, 2018\\
		Michigan State University, East Lansing, Michigan, USA.}

\usepackage{todonotes}
\begin{document}

\section{Introduction}
The current most precise determination of the hadronic vacuum polarization 
(HVP) contribution to the anomalous magnetic moment of the muon $a_\mu$ is 
obtained using the cross section of $e^+e^-\rightarrow$ hadrons (see 
\cite{Keshavarzi:2018mgv,Jegerlehner:2017lbd,Davier:2017zfy} for recent 
results) 
and has an error of about $\lesssim1\%$. A lattice calculation aiming at a 
similar precision requires to include isospin breaking corrections.\par
In nature, isospin symmetry is broken by the different quark masses of the 
up and the down quark and their different electric charges. These effects are 
expected to be of the order $\mathcal{O}((m_d-m_u)/\Lambda_\textrm{\scriptsize 
QCD})$ and $\mathcal{O}(\alpha)$, respectively. In this proceedings we 
present a calculation of isospin breaking corrections to the hadronic vacuum 
polarization at physical quark masses. These results have been published in 
\cite{Blum:2018mom} and are a continuation of our work in \cite{Boyle:2017gzv}, 
where we calculated isospin breaking correction to the HVP at unphysical quark 
masses. Other calculations of isospin breaking corrections to the HVP can be 
found in \cite{Giusti:2017jof,Chakraborty:2017tqp}. \par
The structure of the proceedings is as follows: In section \ref{sec:setup} we 
give details on the computational setup and describe our procedure to tune the 
quark masses to their physical values including isospin breaking corrections. 
In section \ref{sec:res} we discuss our results for QED and strong isospin 
breaking 
corrections. Conclusions and outlook are given in section \ref{sec:conclusions}.

\section{Computational Setup and Tuning of the Quark Masses}
\label{sec:setup}
In this work we calculate isospin breaking corrections using an expansion 
\cite{deDivitiis:2011eh,deDivitiis:2013xla} around the isospin symmetric 
limit, i.e.
\begin{equation}
 C(t) = C^0(t) + \alpha C^{\textnormal{QED}}(t) + \sum_f \Delta m_f C^{\Delta 
m_f}(t) + O(\alpha^2,\alpha\Delta m, \Delta m^2)
\label{eq:expansion}
\end{equation}
for a correlation function $C(t)$, where $C^0(t)$ is the correlation function 
in the isospin symmetric case, $\alpha C^{\textnormal{QED}}(t)$ and $\sum_f 
\Delta m_f C^{\Delta m_f}(t)$ are the leading order QED and strong isospin 
breaking correction, respectively. \par
\begin{figure}[h]
 \centering
 \includegraphics[width=0.98\textwidth]{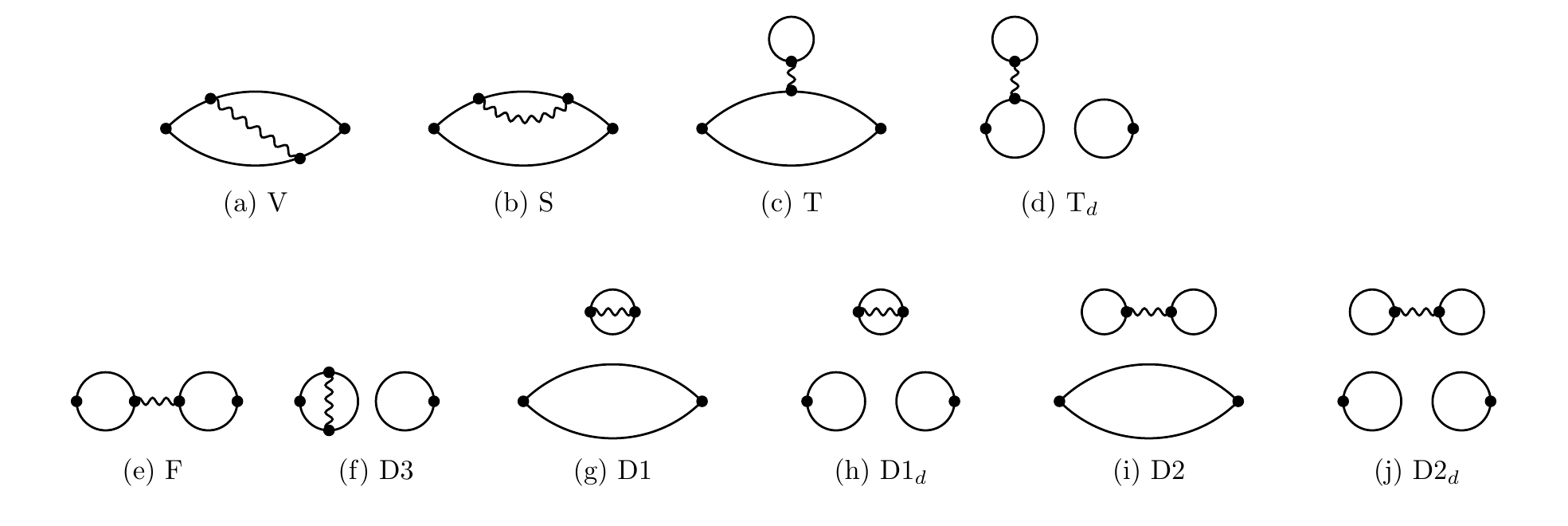}
 \caption{QED correction diagrams at $O(\alpha)$.}
 \label{fig:QEDdiagrams}
\end{figure}

The set of diagrams at $O(\alpha)$ from the 
expansion in the electromagnetic coupling is shown in figure 
\ref{fig:QEDdiagrams}. These can be divided in three different classes of 
diagrams: QED corrections to the quark-connected contribution are given by 
diagrams $V$ and $S$, QED corrections to the quark-disconnected contribution 
are given by diagrams $F$ and $D3$. Diagrams $T$, $D1$, 
$D2$ and $T_d$, $D1_d$, $D2_d$ are
electromagnetic effects for the sea quarks for the quark-connected 
and quark-disconnected contribution, respectively. 
In this work we calculate the 
connected diagrams ($V$, $S$) and the leading disconnected diagram $F$. All 
other diagrams are at least $1/N_c$ or $SU(3)$ flavour suppressed for the HVP 
and 
we add an overall systematic  uncertainty of $30\%$ of the QED correction from 
neglecting these diagrams on our final result for $a_\mu$ in 
\cite{Blum:2018mom}. Note, that we use local vector currents renormalized by 
$Z_V$ for the photon insertions, and thus, tadpole contributions are absent. We 
use Feynman gauge and the QED$_L$ \cite{Hayakawa:2008an} prescription for the 
photon propagators
\begin{equation}
  \Delta_{\mu\nu}(x-y) = 
\delta_{\mu\nu}\,\frac{1}{N}\sum_{k,\vec{k}\neq 0}\,\,
\frac{e^{ik\cdot(x-y)}}{\hat{k}^2}\,.
\label{eq:feynmanqedl}
\end{equation}

\par
\begin{figure}[h]
 \centering
 \includegraphics[width=0.98\textwidth]{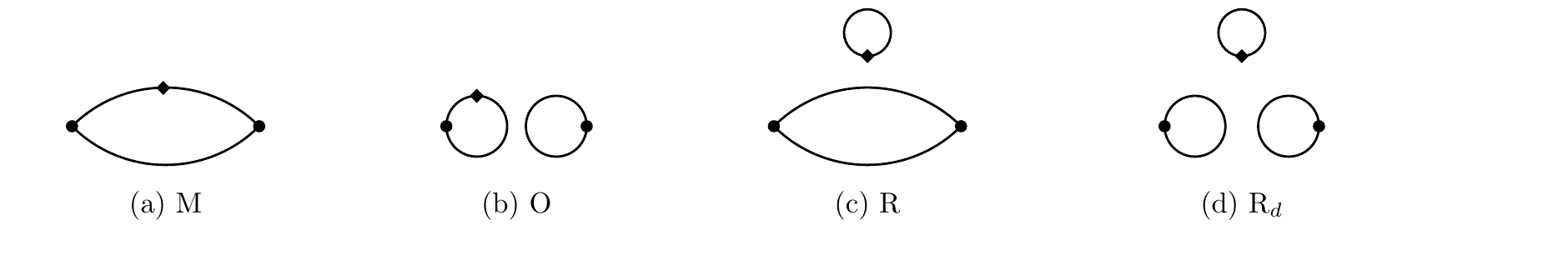}
 \caption{Strong Isospin Breaking correction diagrams at $O(\Delta m)$.}
 \label{fig:sIBdiagrams}
\end{figure}

Isospin breaking corrections due to the expansion in the quark masses in 
equation (\ref{eq:expansion}) are given by the diagrams in figure 
\ref{fig:sIBdiagrams}. In this work, we only calculate the quark-connected 
correction (diagram $M$). We neglect mass corrections to the quark-disconnected 
contribution (diagram $O$), which are $1/N_c$ and $SU(3)$ flavour suppressed 
and 
assign an additional $10\%$ of the strong isospin breaking correction as an 
systematic error in the final result for $a_\mu$ \cite{Blum:2018mom}. The 
mass correction to the sea quarks (diagram $R$ and $R_d$ for the 
quark-connected and quark-disconnected HVP, respectively) is proportional to a 
factor of 
($\Delta m_u + \Delta m_d)$ and we find $\Delta m_u\approx - \Delta m_d$ when 
tuning the quark masses to their physical values and thus expect diagram $R$ to 
be negligible.

\par
We calculate the isospin breaking corrections to the HVP on a $48^3\times96$ 
lattice using $N_f=2+1$ dynamical flavours of Mobius Domain Wall Fermions on a 
single lattice spacing $a^{-1}=1.730(4)$~GeV. The isospin symmetric calculation 
is done using a light and a strange quark mass, that have been tuned to 
reproduce a pion of $m_\pi^0=135.0$~MeV and a kaon of $m_K^0=495.7$~MeV in the 
absence of QED and strong isospin breaking 
effects~\cite{Blum:2014tka}. To obtain up, down and strange quark
masses at their physical values including QED, we proceed as follows. We fix 
the 
charged pion, neutral kaon and charged kaon masses including QED to their 
experimental values
 \begin{equation}
 a\,m_{\pi^+}^\textrm{exp} = \left[m^0_{\pi} + 
 \alpha m^\textnormal{QED}_{\pi^+} + \Delta 
 m_d \,\, m^{\Delta m_d}_{\pi^+} +\Delta m_u\, \, m^{\Delta m_u}_{\pi^+} 
 \right]
 \label{eq:pi+mass}
 \end{equation}
  \begin{equation}
a \, m_{K^+}^\textrm{exp} = \left[m^0_{K}  + 
\alpha m^\textnormal{QED}_{K^+} + \Delta 
m_u\,\,m^{\Delta m_u}_{K^+} + \Delta 
m_s\,\,m^{\Delta m_s}_{K^+} \right]
 \label{eq:K+mass}
 \end{equation}
 \begin{equation}
a\, m_{K^0}^\textrm{exp} = \left[ m^0_{K}  + 
\alpha m^\textnormal{QED}_{K^0} + \Delta 
m_d\,\,m^{\Delta m_d}_{K^0} + \Delta 
m_s\,\,m^{\Delta m_s}_{K^0}   \right]
 \label{eq:K0mass}
\end{equation}
where $m^0_{H}$ is the isospin symmetric mass of $H$, $\alpha 
m^\textnormal{QED}_H$ the QED correction to mass of $H$ and $\Delta m_f\,\, 
m^{\Delta m_f}_{H}$ the correction to the mass of $H$ from a shift of the 
quark mass 
$\Delta m_f = (m_f-m_f^0)$. The shifts $\Delta m_f$ in the quark masses are 
free parameters that can be tuned after all the required correlation functions 
have been calculated. Once 
having tuned the quark masses to reproduce the physical values of $\pi^+$, 
$K^+$ and $K^0$ we checked that we also correctly reproduce the neutral pion 
mass. \par
In addition, the tuning of the quark masses as in equations 
(\ref{eq:pi+mass}) - (\ref{eq:K0mass}) requires to determine the lattice 
spacing in the presence of QED. Here, we choose to set the lattice spacing by 
fixing the mass of the $\Omega^-$ baryon 
\begin{equation}
a \rightarrow a(\Delta m_s) = \left(m^0_\Omega +  
\alpha m^\textnormal{QED}_\Omega + 3\,\Delta 
m_s\,\,m^{\Delta m_s}_{\Omega}\right)/m^\textnormal{exp}_\Omega\,.
\end{equation}
We find the shift in the lattice 
spacing to be smaller then the statistical error on the lattice 
spacing and therefore neglect this effect in the following.

\section{Results}
\label{sec:res}
In the following we discuss our results for different contributions of the QED 
and 
strong isospin breaking corrections to the hadronic vacuum polarization. The 
HVP contribution to the anomalous magnetic moment of the muon can be calculated 
from the vector-vector two-point function~\cite{Bernecker:2011gh,Feng:2013xsa}
\begin{equation}
 a_\mu = \sum_t w_t C(t)\hspace{1.4cm}\textnormal{with}\qquad C(t) = 
\frac{1}{3}\sum_{j=0}^2\sum_{\vec{x}} \left<J_j(\vec{x},t)J_j(0)\right>
\label{eq:amu}
\end{equation}
where $J_j = \frac{2}{3}\overline{u}\gamma_j u - 
\frac{1}{3}\overline{d}\gamma_j d + \cdots$ are electromagnetic vector 
currents. In this work we use local vector currents multiplied with the 
vector current renormalization $Z_V$. We 
also calculate the QED correction to $Z_V$ and find this to be negligible for 
our setup.
\subsection{Quark-connected QED correction}
\begin{figure}[h]
 \centering
 \includegraphics[width=0.5\textwidth]{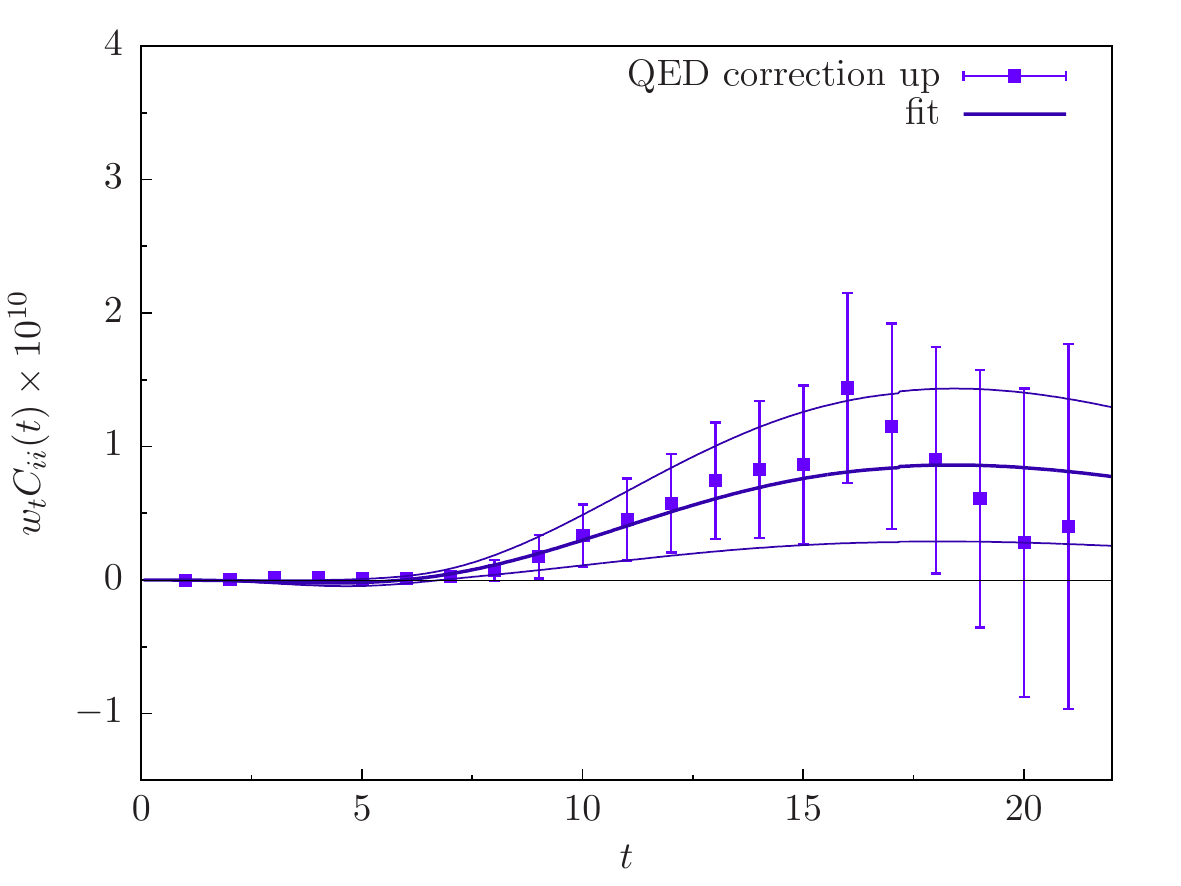}
 \caption{Results for the quark-connected QED correction to $w_t C(t)$. The 
result of the fit is shown by the solid line, with the fainter lines indicating 
the error band of the fit.}
\label{fig:qedcon}
\end{figure}

Our data for the quark-connected QED correction (diagrams $V$ and $S$ in figure 
\ref{fig:QEDdiagrams}) to the integrand $w_t C(t)$ is shown in figure 
\ref{fig:qedcon}. The QED correction to $a_\mu$ from these contributions can 
then be 
obtained by integrating the data over the euclidean time $t$. However, as 
one can see in figure \ref{fig:qedcon} the statistical error on the data is
large. Therefore, we replaced the data in the 
integration for $a_\mu$ by a fit ansatz
\begin{equation}
 C(t) = (c_1 + c_0\, t) e^{-Et}\,.
 \label{eq:fitansatz}
\end{equation}
We fix $E$ to the energy of the lowest lying state, which, including QED, is 
given by $\pi\gamma$, where in QED$_L$ \cite{Hayakawa:2008an} the photon has 
one unit of momentum (cf. equation (\ref{eq:feynmanqedl})). We then fit our 
data to the 
ansatz (\ref{eq:fitansatz}) using $c_1$ and $c_2$ as free parameters. The 
result of this fit is shown in figure \ref{fig:qedcon} by the solid line.\par
For the connected QED correction to $a_\mu$ we find
\begin{equation}
  a_\mu^{\textnormal{QED,con}} = 
5.9(5.7)_S(1.1)_E(0.3)_C(1.2)_V(0.0)_A(0.0)_Z\times 10^{-10}\,,
\label{eq:cQEDres}
\end{equation}
where the first error $()_S$ is statistical. Further to that, we have assigned 
the following systematic errors. A systematic error from our fit ansatz $()_E$ 
is
determined by varying the input for the lowest energy between $\pi\gamma$ and 
$\pi\pi$. We estimate a discretization error $()_C$ as $(a\Lambda)^2$ with 
$\Lambda=400$~MeV. When including QED in a lattice 
calculation, finite volume corrections can be large (see e.g.\ 
\cite{Borsanyi:2014jba}). 
In this work, we estimate finite volume corrections, be replacing the photon 
propagator by its infinite volume expression and take the difference in the 
final result as a systematic error $()_V$ from not correcting for these 
effects. More details on using an infinite volume photon propagator can be 
found in the supplementary material of \cite{Blum:2018mom}.
However, a study presented at this conference \cite{harrisonproc} suggests, 
that the finite volume effects for the QED correction to the HVP are much 
smaller. Finally, we propagate uncertainties from the lattice spacing $()_A$ 
and vector renormalization constant $()_Z$, but find them to be negligible 
compared to other sources of systematic errors.
\subsection{Quark-disconnected QED correction}
The leading QED correction to the quark-disconnected HVP is given by diagram 
$F$ in figure~\ref{fig:QEDdiagrams}. Here, we are only interested in 
contributions, where in addition to the photon the quark lines are connected by 
gluons. If no additional gluons connect both quark lines, these 
contributions are conventionally counted as higher order HVP contributions and 
need to be subtracted in this context.\par
We calculated the leading quark-disconnected QED correction to the HVP using 
data generated for the light-by-light scattering project \cite{Blum:2016lnc}. 
The data is fitted using the same ansatz (\ref{eq:fitansatz}) as for the 
connected QED correction. We 
find for the leading quark-disconnected QED correction to the anomalous 
magnetic 
moment
\begin{equation}
 a_\mu^{\textnormal{QED, disc}} =  
-6.9(2.1)_S(1.3)_E(0.4)_C(0.4)_V(0.0)_A(0.0)_Z\times 10^{-10}\,,
\label{eq:dQEDres}
\end{equation}
where we have assigned a similar set of systematic errors as above.
\subsection{Strong Isospin breaking corrections}
\begin{figure}[h]
 \centering
 \includegraphics[width=0.5\textwidth]{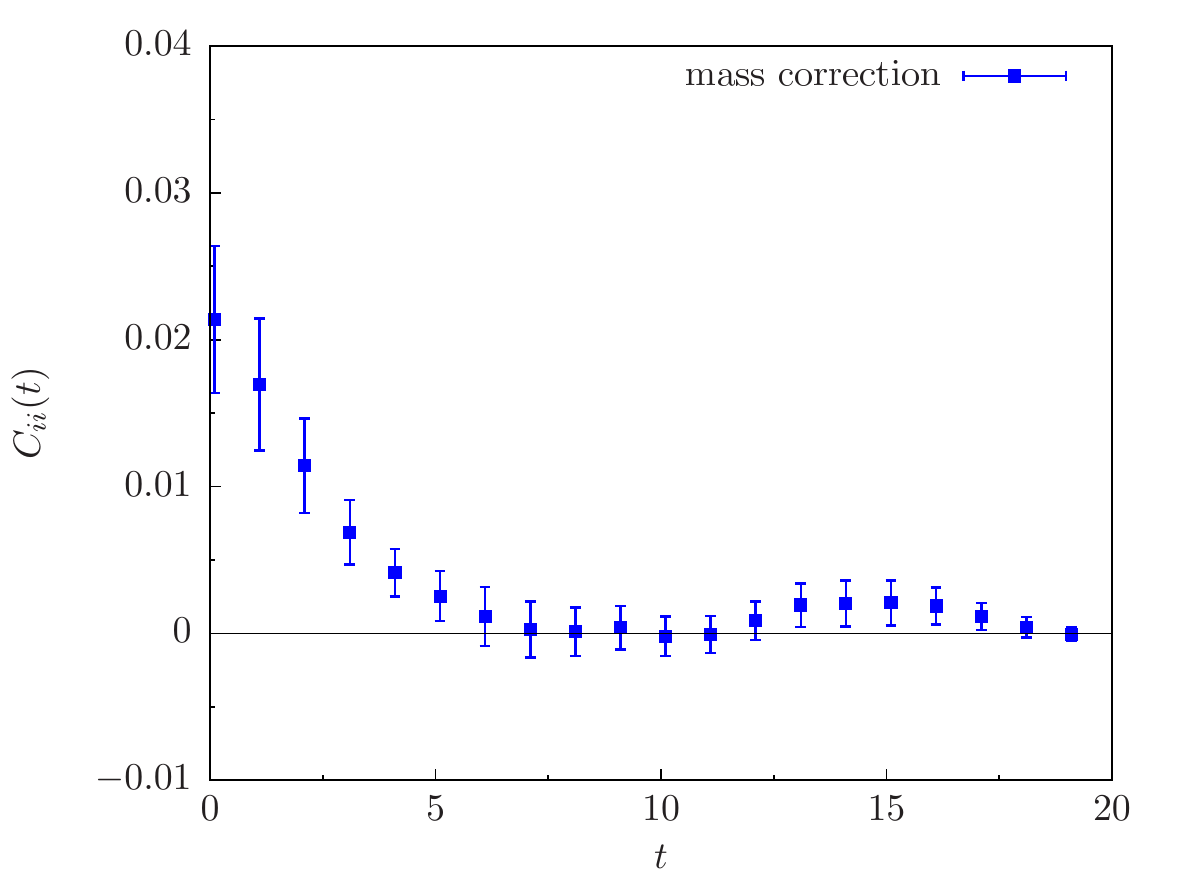}
 \caption{The isospin breaking correction to $C(t)$ from the 
connected mass insertion diagram $M$.}
\label{fig:CsIB}
\end{figure}

In figure \ref{fig:CsIB} we show the correction to the vector two-point 
function $C(t)$ for the quark-connected mass insertion diagram (diagram $M$ in 
figure \ref{fig:sIBdiagrams}). We fit the correlator using a similar ansatz 
(\ref{eq:fitansatz}) as for QED, using $\pi\pi$ as the lowest lying energy 
state as an input. Here, we vary the energy between two interacting and two 
free pions to obtain a systematic error $()_E$ from the fit ansatz.
\par
For the correction to $a_\mu$ from diagram $M$ we find
\begin{equation}
 a_\mu^{\textnormal{sIB}} = 
10.6(4.3)_S(1.3)_E(0.6)_C(6.6)_V(0.1)_A(0.0)_Z\times 10^{-10}\,.
\label{eq:sIBres}
\end{equation}
For the strong isospin breaking correction we estimate finite volume 
corrections using chiral perturbation theory 
\cite{Bijnens:2017esv}.

\section{Conclusions}
\label{sec:conclusions}
The results shown here are 
part of RBC/UKQCD's recently published result for the anomalous magnetic 
moment of the muon \cite{Blum:2018mom}. This work is the first calculation of 
the HVP contribution to the $a_\mu$ including a 
calculation of QED and strong isospin breaking corrections directly at the 
physical point. Although our statistical errors are 
still large, we find isospin breaking corrections to be at the order of $1\%$, 
with results for the connected QED correction given in (\ref{eq:cQEDres}), the 
leading disconnected QED correction given in (\ref{eq:dQEDres}) and the 
connected strong isospin breaking correction given in (\ref{eq:sIBres}). For 
the future, we plan to increase the statistics on these contributions in 
addition to calculating the missing quark-disconnected diagrams and studying 
the effect of isospin breaking corrections for the sea quarks. In that context 
we are looking into reusing the data from the hadronic light-by-light 
scattering project \cite{Blum:2016lnc} for all of the diagrams shown in figure 
\ref{fig:QEDdiagrams}. Further to 
that, we will extend our study of isospin breaking corrections to a second 
lattice spacing in oder to take a continuum limit. 
\par
These future improvements will 
allow us to determine the isospin breaking correction to $a_\mu^{HVP}$ precise 
enough to be consitent with reaching an overall error of $<1\%$ on the total 
QCD$+$QED result for $a_\mu^{HVP}$ from a lattice calculation.

\section*{Acknowledgments}
V.G. and A.P. received funding from the European Research Council (ERC) under 
the European Union's Horizon 2020 research and innovation programme under grant 
agreement No 757646. A.P also received funding from UK STFC grants ST/L000458/1 
and ST/P000630/1.
C.L. is supported by US DOE Contract DESC0012704(BNL) and by a DOE Office of 
Science Early CareerAward.
V.G. and A.J. received funding from STFC consolidated grant 
ST/P000711/1. A.J. received funding from the European
618 Research Council under the European Union's Seventh Framework Program 
(FP7/2007-619 2013) / ERC Grant agreement 279757. 
This work was supported by resources provided by the Scientific Data and 
Computing  Center (SDCC) at Brookhaven National Laboratory (BNL), a DOE 
Office of Science User Facility supported by the Office of Science of 
the US Department of Energy. The SDCC is a major component of the Computational 
Science Initiative at BNL. We gratefully acknowledge computing resources 
provided through USQCD clusters at Fermilab and Jefferson Lab. This work was 
also supported by the DiRAC Blue Gene Q Shared Petaflop system at the University 
of Edinburgh, operated by the Edinburgh Parallel Computing Centre on behalf of 
the STFC DiRAC HPC Facility (www.dirac.ac.uk). This equipment was funded by BIS 
National E-infrastructure capital grant ST/K000411/1, STFC capital grant 
ST/H008845/1, and STFC DiRAC Operations grants ST/K005804/1 and ST/K005790/1. 
DiRAC is part of the UK National E-Infrastructure.

\end{document}